\documentclass[journal,letterpaper,twoside,twocolumn]{IEEEtran}
\usepackage{amsmath,amssymb,graphicx,psfrag,cite}
\usepackage[usenames,dvipsnames]{color} %
\usepackage{balance}
\usepackage{mathrsfs}   %

\newcommand{\bc}{{\boldsymbol{c}}}
\newcommand{\bw}{{\boldsymbol{w}}}
\newcommand{\bx}{{\boldsymbol{x}}}
\newcommand{\by}{{\boldsymbol{y}}}
\newcommand{\bR}{{\boldsymbol{R}}}
\newcommand{\bX}{{\boldsymbol{X}}}
\newcommand{\bY}{{\boldsymbol{Y}}}
\newcommand{\bZ}{{\boldsymbol{Z}}}
\newcommand{\bbeta}{{\boldsymbol{\beta}}}
\newcommand{\bDelta}{{\boldsymbol{\Delta}}}
\newcommand{\cC}{{\mathscr{C}}}
\newcommand{\C}{{\mathcal{C}}}
\newcommand{\Ci}{C_\text{i}}
\newcommand{\E}{\mathbb{E}}
\newcommand{\X}{\mathcal{X}}
\newcommand{\Y}{\mathcal{Y}}
\newcommand{\R}{\mathbb{R}}
\newcommand{\amax}{{a_\mathrm{max}}}
\newcommand{\p}{\mathsf{P}} %

\newcommand{\eqlab}[2]{\begin{align} \label{#1} #2 \end{align}}
\newcommand{\eq}[1]{\begin{align} #1 \end{align}} %

\newcommand{\disable}[1]{}

\newcommand{\new}[1]{{#1}}

\newtheorem{theorem}{Theorem}
\newtheorem{lemma}[theorem]{Lemma}
\newtheorem{corollary}[theorem]{Corollary}

\newtheorem{definition}{Definition}

\title{Conditions for a Monotonic Channel Capacity}
\author{Erik Agrell, \IEEEmembership{Senior Member, IEEE}
\thanks{This work is an expanded version of ``The channel capacity increases with power,'' http://arxiv.org/abs/1108.0391, 2011. It was supported in part by the Swedish Foundation for Strategic Research (SSF) under grant RE07-0026 and the Swedish Research Council (VR) under grant 2007-6223. E.~Agrell is with the Dept.~of Signals and Systems, Chalmers Univ.~of Technology, SE-41296 G\"oteborg, Sweden (e-mail: agrell@chalmers.se).}}

\begin{document}
\maketitle

\begin{abstract}
Motivated by results in optical communications, where the performance can degrade dramatically if the transmit power is sufficiently increased, the channel capacity is characterized for various kinds of memoryless vector channels. It is proved that for all {static} point-to-point channels, the channel capacity is a nondecreasing function of power. As a consequence, maximizing the mutual information over all input distributions with a certain power is for such channels equivalent to maximizing it over the larger set of input distributions with upperbounded power. For interference channels such as optical wavelength-division multiplexing systems, the primary channel capacity is always nondecreasing with power if all interferers transmit with identical distributions as the primary user. Also, if all input distributions in an interference channel are optimized jointly, then the achievable sum-rate capacity is again nondecreasing. The results generalizes to the channel capacity as a function of a wide class of costs, not only power.
\end{abstract}

\section{Introduction} \label{sec:intro}

\IEEEPARstart{I}{n the most} cited paper in the history of information theory \cite{shannon48}, Shannon in 1948 proved that with adequate coding, reliable communication is possible over a noisy channel, as long as the rate does not exceed a certain threshold, called the \emph{channel capacity.} He provided a mathematical expression for the channel capacity of any point-to-point channel, based on its statistical properties. The expression is given as the supremum over all possible input distributions of a quantity later called the \emph{mutual information} \cite{shannon56, kolmogorov56}. The channel capacity is often studied as a function of a cost, such as the transmit power. More specifically, the capacity--cost function is defined as the supremum of the mutual information over all input distributions whose cost is either \emph{equal} to a given constant or \emph{upperbounded} by a constant---the convention differs between disciplines. We will return to the distinction between the two definitions at the end of this section.

For the \emph{additive white Gaussian noise} (AWGN) channel, the channel capacity is known exactly \cite[Sec.~24]{shannon48}, \cite[Ch.~9]{cover06}. In recent years, the problem of calculating or estimating the channel capacity of more complicated channels has received a lot of attention (see surveys in \cite{katz04, kahn04, essiambre10, killey11}).  Due to the absence of exact analytical solutions and the computational intractability of optimizing over all possible input distributions, most investigations of the channel capacity of non-AWGN channels rely on bounding techniques and asymptotic analysis.

The main motivation for this paper comes from the type of nonlinear distortion encountered in fiber-optical communications. In contrast to linear channels, an optical fiber has the peculiar property that \new{the performance of conventional communication systems degrades if the signal amplitude is increased beyond a certain level} \cite{mecozzi94, demir07, bononi12, carena12, beygi12, agrell12, hager12}, \cite[Ch.~9]{agrawal10}. This phenomenon is well-known from experiments and simulations, and can also be explained theoretically. The lightwave propagation in an optical fiber is governed by a nonlinear differential equation, the \emph{nonlinear Schr\"odinger equation} or, if polarization effects are considered, the \emph{Manakov equation} \cite{chraplyvy90, wang99}, \cite[Sec.~7.2]{agrawal08book}. These equations include a nonlinear distortion term, whose amplitude is proportional to the cubed signal amplitude. At high enough signal amplitudes, this nonlinear distortion dominates the other terms in the differential equation, effectively drowning the signal.

Similarly, one might expect that the nonlinear distortion would force the mutual information and channel capacity down to zero at sufficiently high power, and in the past two decades, many results have been published in optical communications to support this conjecture {\cite{mitra01, stark01, tang01, green02, narimanov02, wegener04, kahn04, ho05, djordjevic05, tang06, taghavi06, essiambre08, freckmann09, essiambre10, djordjevic10, ellis10, smith10, goebel11, killey11, bosco11, mecozzi12}}. Already in 1993, Splett \emph{et al.} modeled the interference from four-wave mixing in a wavelength-division multiplexing (WDM) system as an AWGN component, under some conditions on the noise and dispersion, in what might have been the first study ever of the channel capacity of a nonlinear optical link \cite{splett93}. The variance of this AWGN depends nonlinearly on the transmit power, which is assumed equal on all wavelengths. Similar nonlinear channel models have been rediscovered, modified, and further analyzed in \cite{stark01, wegener04, poggiolini11, bononi12, bosco11, carena12, beygi12, poggiolini12, poggiolini14}. Due to the signal-dependent noise, their channel capacities are not monotonic: As the transmit power (or signal-to-noise ratio) increases, the channel capacity increases towards a peak and then decreases again as the power is further increased. Other channel models with signal-dependent AWGN were presented in \cite{goebel11, mecozzi12} and have similar nonmonotonic channel capacities. An essential assumption, explicit or implicit, in the derivation of these AWGN-based models is that the transmitted signal consists of independent, identically distributed symbols. This assumption is valid in uncoded transmission systems, but not in the presence of {error-correction} coding, since coding introduces correlation between symbols. Using a model derived under certain conditions on the transmitted signal is particularly risky in channel capacity calculations, since the channel capacity is by definition the maximum achievable rate using \emph{any} transmission scheme---including those for which the constrained model is not valid.

A continuous-time channel model for cross-phase modulation (XPM) was presented by Mitra and Stark \cite{mitra01}. Although no discrete-time XPM model was obtained, they showed that the channel capacity of the XPM channel model is lowerbounded by the capacity of a signal-dependent AWGN channel, and that this lower bound is nonmonotonic. They further conjectured that the true channel capacity would have a similar nonmonotonic behavior as its lower bound. Many variants of the Mitra--Stark lower bound have been presented in recent years, often along with the conjecture that the true channel capacity is also nonmonotonic \cite{green02, wegener04, tang06, ellis10, ellis11}. {This conjecture was disproved in the zero-dispersion case by Turitsyn \emph{et al.} \cite{turitsyn03}, who showed that the lower bound based on the AWGN channel \cite{tang01} is very far from the true channel capacity,} and that the channel capacity in fact grows logarithmically with power under certain conditions.

Another type of lower bound on channel capacity is obtained by fixing the input distribution and calculating the mutual information \cite{ho02, djordjevic05, essiambre10, smith10, djordjevic10, djordjevic11} or by optimizing the mutual information over a subset of all possible input distributions \cite{essiambre08, freckmann09, essiambre10, smith10, goebel11}. All these lower bounds consistently show a nonmonotonic behavior, decreasing towards zero after a peak at a finite power, and the conjecture that the channel capacity would have a similar nonmonotonic behavior as its lower bounds is often repeated.

We believe that the results cited above, while mathematically correct, do not fully exploit the potential of capacity-achieving coding over nonlinear optical channels. {We prove in this work mathematically that for a wide class of channel models,} the capacity is a monotonic function (nondecreasing but not necessarily strictly increasing) {of the transmit power. This property holds for any \emph{static} channel model, defined as one whose channel law does not change depending on which input distribution it is combined with.} The results are extended to a wide class of cost functions and to three specific multiuser scenarios.

The presented \new{results} holds regardless of whether the capacity--cost function is defined by maximizing over all input distributions with exactly the given cost or with an upperbounded cost. The proofs are developed assuming the former definition, and they are all trivial for the latter. An interesting consequence of the nondecreasing channel capacity is that the two definitions of the capacity--cost function are fully equivalent.

\section{Channel Capacity {and Cost}} \label{sec:def}%

Let $\bX$ and $\bY$ be real, $n$-dimensional vectors, representing the input and output, resp., of a discrete-time memoryless communication channel. {Their respective domains, or alphabets, are denoted by $\X\subseteq\R^n$ and $\Y\subseteq\R^n$.} The joint distribution $f_{\bX,\bY}(\bx,\by)$ for $\bx \in \X$ and $\by\in\Y$ can be factorized as $f_{\bX,\bY}(\bx,\by) = f_\bX(\bx) f_{\bY|\bX}(\by|\bx)$, where $f_\bX$ {is the \emph{input distribution}} (which is in practice determined by the modulation format) and $f_{\bY|\bX}$ {is the \emph{channel law.}} We denote the \emph{mutual information} between $\bX$ and $\bY$ with $I(\bX;\bY)$, while $I(\bX;\bY|\bZ)$ denotes a \emph{conditional mutual information.} The \emph{entropy} and \emph{conditional entropy} are denoted by $H(\bX)$ and $H(\bX|\bZ)$, resp., and the \emph{differential entropy} and \emph{conditional differential entropy} are denoted by $h(\bX)$ and $h(\bX|\bZ)$, resp.

Using {error-correction} coding, codewords of \new{$N$} input symbols are selected from a codebook \new{$\C \subseteq \X^N$}. The \emph{rate} of transmission, in bits per \new{symbol}, is \new{$\log_2|\C|/N$}. The codewords can be transmitted with arbitrarily small error probability if the codewords are sufficiently long and the rate is sufficiently small. Such a rate is called an \emph{achievable rate,} and the supremum of all achievable rates, over all possible codes and block lengths, is defined as the \emph{operational channel capacity} \new{or simply \emph{capacity}Ê\cite[Sec.~1, 14]{shannon48}.}

Shannon's channel coding theorem \cite[Sec.~13, 23]{shannon48}, \cite[Sec.~7.7, 9.1]{cover06} states that the operational channel capacity is equal to the \emph{information channel capacity}, which is defined as the supremum of the mutual information $I(\bX;\bY)$ between the channel input and output, where the supremum is taken over all input distributions $f_\bX$. The capacity-achieving distribution may be continuous or discrete \cite{smith71, gursoy05}{.}%
\footnote{With a slight abuse of notation, we also include distributions that have no {probability density function} \cite[Sec.~8.5]{cover06}.}

In this work, the channel capacity is characterized as a function of some kind of cost. Closely following the definitions in \cite{verdu90}, \cite[Ch.~2]{mceliece02}, \cite[Sec.~3.3]{elgamal11}, we define the \emph{cost function} $b(\bx)$ as a deterministic, real, nonnegative function of an input symbol $\bx\in\X$.
\new{The cost of a codeword $\bc = (\bx_1,\ldots,\bx_N)$ is defined as $b(\bc) = (b(\bx_1)+\cdots+b(\bx_N))/N$. Let $\tilde{M}(N,p,\beta)$ be the size of the largest codebook $\tilde{\C}$ such that (i) $b(\bc)\le\beta$ for all $\bc\in\tilde{\C}$ and (ii) each codeword can be decoded with an error probability not larger than $p$. The \emph{capacity--cost function} is defined as \cite[Sec.~7.5, 9]{cover06} \cite[Sec.~3.3]{elgamal11}
\eqlab{eq:cprimedef}{
\tilde{C}(\beta) \triangleq \lim_{p\rightarrow0}\lim_{N\rightarrow\infty} \frac{\log_2 \tilde{M}(N,p,\beta)}{N}
.}
The channel coding theorem with an upperbounded cost now states that \cite[Sec.~7.3]{gallager68}, \cite[Sec.~3.3]{elgamal11}
\eqlab{eq:cprime}{
  \tilde{C}(\beta) = \sup_{f_\bX \in \tilde{\Omega}(\beta)} I(\bX;\bY)
,}
where $\tilde{\Omega}(\beta)$ is the set of all distributions $f_\bX$ over $\X$ such that $\E[b(\bX)] \le \beta$. It is well known that the channel capacity, as defined above, is nondecreasing with $\beta$ \cite{taghavi06}, \cite[Ch.~2]{mceliece02}, \cite[Sec.~3.3]{elgamal11}. This follows from \eqref{eq:cprime} and the fact that $\tilde{\Omega}(\beta) \supseteq \tilde{\Omega}(\beta')$ for all $\beta \ge \beta'$. 

In this paper, we focus on another type of cost constraint. Instead of upperbounding the cost of the codewords as in the previous paragraph, the codewords are all required to have the same exact cost. This scenario has been touched upon in the past \cite{shannon59,polyanskiy10}, but not received as rigorous information-theoretic treatment as the bounded-cost constraint. Formally, let $M(N,p,\beta)$ be the size of the largest codebook $\C$ such that (i) $b(\bc)=\beta$ for all $\bc\in\C$ and (ii) each codeword can be decoded with an error probability not larger than $p$. In analogy with \eqref{eq:cprimedef}, the capacity--cost function is defined as
\eqlab{eq:cdef}{
C(\beta) \triangleq \lim_{p\rightarrow0}\lim_{N\rightarrow\infty} \frac{\log_2 M(N,p,\beta)}{N}
.}
It is also possible to define the information capacity with an equality constraint, analogous to the right-hand side of \eqref{eq:cprime}, as
\eqlab{eq:c}{
  \Ci(\beta) \triangleq \sup_{f_\bX \in \Omega(\beta)} I(\bX;\bY)
,}
where $\Omega(\beta)$ is the set of all distributions $f_\bX$ over $\X$ such that $\E[b(\bX)] = \beta$. This quantity has been analyzed and characterized extensively in optical communications (e.g., \cite{mitra01,wegener04,essiambre08},{\cite[eq.~(11.5)]{ho05}}) and also considered in wireless communications \cite{taricco97}, which partly motivates this work. The interest in $\Ci(\beta)$ comes from an implicit assumption that a channel coding theorem would exist also with an equal-cost constraint, i.e., that $C(\beta) = \Ci(\beta)$. This relation, while intuitively reasonable, was to our knowledge never formally proven, which may cast some doubts on the operational interpretation of any results based on $\Ci(\beta)$. In the next section, it will be shown that $\Ci(\beta)$ is nondecreasing with $\beta$ and, as a consequence thereof, that indeed $C(\beta) = \Ci(\beta)$.
}

An implicit assumption for \eqref{eq:c} is that $I(\bX;\bY)$ is calculated from the same channel law $f_{\bY|\bX}$ for all $f_\bX \in \Omega(\beta)$ and all $\beta>0$. In other words, the channel remains the same regardless of which codebook is used.
This is a standard assumption in information theory \cite[Sec.~4.2]{gallager68}, and it is not considered to restrict generality. Channel laws with this property are formally defined as \emph{static} in the next section.

Readers with an information theory background have probably only encountered static channel models and may not see the need to define a name for channels with this property. In \new{the} optical communications \new{literature}, however, channel models $f_{\bY|\bX}$ that change with $f_\bX$ \new{have been proposed frequently}. Consider for example the well-known Gaussian noise model for fiber-optical links without dispersion compensation \cite{splett93, poggiolini11, bosco11, bononi12, carena12, beygi12, mecozzi12, poggiolini12, poggiolini14}. In its simplest form, the model is given by the channel law
\eqlab{gn}{
f_{Y|X}(y|x) = \frac{1}{\pi(\sigma_0^2+\eta \p^3)} e^{-\frac{|y-x|^2}{\sigma_0^2+\eta \p^3}}
,}
where $X$ and $Y$ are the complex channel input and output, resp., $\sigma_0^2$ and $\eta$ are two constant link parameters, and $\p = \E[|X|^2]$. This is an AWGN channel, whose noise variance depends on $\p$ and hence on $f_\bX$.\footnote{\new{A \emph{static} model for a similar channel as \eqref{gn} was given in \cite[Eq.~(13)]{agrell14jlt}.}}
Its \new{information} capacity, obtained by Shannon's standard formula \cite[Sec.~24]{shannon48}, \cite[Sec.~9.1]{cover06}, is commonly given as \cite{splett93, bosco11, poggiolini14}
\eqlab{cgn}{
C(\p) = \log_2\left(1+\frac{\p}{\sigma_0^2+\eta \p^3} \right)
.}
This function, which clearly decreases to zero at high power $\p$, exemplifies the nonmonotonic behavior of the capacity of certain optical (nonstatic) channel models. We advise that such channel models, while unarguably accurate in some scenarios \cite{carena12, poggiolini14}, should be used with caution in information-theoretic analysis. First, it is not clear whether \eqref{eq:c} has any operational meaning in terms of maximum achievable rates for nonstatic channels \new{such as \eqref{gn}}. Shannon's channel coding theorem, in its standard memoryless form, assumes that the channel law operates on each symbol $\bX$ independently, which is not the case if $f_{\bY|\bX}$ changes with $f_\bX$.
And second, such models are questionable from a physical viewpoint, as they imply an infinite channel memory \cite{agrell13ecoc}, \cite{agrell14jlt}.
Only static channel models will be considered \new{further in this paper}.

\section{Point-to-Point Channels}

In this section, we are concerned with a discrete-time, memoryless vector channel between a single transmitter and a single receiver, formally defined as follows.

\begin{definition}\label{def:static}
A {\emph{static}} point-to-point channel is a memoryless relationship $f_{\bY|\bX}{(\by|\bx)}$ between vectors $\bX \in \X$ and $\bY \in \Y$, which {is a function of $\by$ and $\bx$ but does not change with $f_\bX$.}
\end{definition}

Such a relationship can represent a continuous-time bandlimited channel by sampling the transmitted and received waveforms at the Nyquist rate \cite[Sec.~23]{shannon48}, and it can {represent channels with an arbitrarily long (finite) memory by choosing the dimension $n$ much larger than the channel memory \cite[Sec.~4.6]{gallager68}, \cite{verdu94}. The dimensions may also, in addition to time, represent frequency (wavelength), space, polarization, lightwave modes, or all of these. Hence, the theory applies to a wide variety of channels in different applications.}

The capacity is commonly studied as a function of the transmit power, which is obtained by setting $b(\bx) = \|\bx\|^2$ for all $\bx \in \X = \R^n$. The results in this paper hold not only for transmit power but also more generally for any unbounded cost function, according to the following definition.

\begin{definition}
An \emph{unbounded cost function} $b(\bx)$ over a domain $\X$ is a real, nonnegative function such that for any given $b_0\ge 0$, there exists a vector $\bx\in\X$ for which $b(\bx) = b_0$.
\end{definition}

The main result for point-to-point channels is the following theorem, which implies that the channel capacity will either increase indefinitely or converge to a finite value as the cost increases, depending on the channel. However, it cannot have a peak for any channel or any cost. Despite its simple nature, it has to our knowledge not been stated before.

\begin{theorem}[Monotonic Channel Capacity]\label{th:main}
{Let $f_{\bY|\bX}(\by|\bx)$ be a {static} point-to-point channel defined on $\bx\in\X$ and $\by\in\Y$. Let $b(\bx)$ be an unbounded cost function on $\X$. Then $\new{\Ci}(\beta)$ is a nondecreasing function of $\beta$.}
\end{theorem}

\begin{IEEEproof}
We will show that for any given pair of costs $\beta \ge \beta' \ge 0$, $\new{\Ci}(\beta) \ge \new{\Ci}(\beta')$. Let, for any $0<\epsilon\le 1$,
\eq{
\beta'' \triangleq \beta'+\frac{\beta-\beta'}{\epsilon}
.}
We define a time-sharing random symbol $\bX {\in \X}$ given an auxiliary binary random variable $Q$ such that
\eq{
  \bX \triangleq \begin{cases}
    \bX', & Q = 0, \\
    \bX'', & Q = 1,
  \end{cases}
}
where $\Pr\{Q = 1\} = \epsilon$ and the distributions of $\bX'{\in \X}$ and $\bX''{\in \X}$ satisfy $\E[b(\bX')] = \beta'$ and $\E[b(\bX'')] = \beta''$, resp. {Such distributions exist, by assumption, for any costs $\beta', \beta'' \ge 0$}. Thus
\eq{
  \E[b(\bX)] &= (1-\epsilon)\E[b(\bX')]+\epsilon \E[b(\bX'')] \notag\\
  & = (1-\epsilon)\beta' + \epsilon \beta'' \notag\\
  & = \beta
  .}

Because $Q\rightarrow\bX\rightarrow\bY$ is a Markov chain, the mutual information can be bounded as
\eqlab{ixy-lb}{
I(\bX{;}\bY) &\ge I(\bX;\bY | Q) \notag\\
  &= (1-\epsilon) I(\bX;\bY | Q=0) + \epsilon I(\bX;\bY | Q=1) \notag\\
  &\ge (1-\epsilon) I(\bX;\bY | Q=0) \notag\\
  &= (1-\epsilon) I(\bX';\bY')
  ,}
where the first inequality follows from{\cite[eq.~(2.122)]{cover06}} and $\bY'$ is defined as the channel output when the input is $\bX'$. {This inequality holds for any $0<\epsilon\le 1$ and any distributions $f_{\bX'} \in \Omega(\beta')$ and $f_{\bX''} \in \Omega(\beta'')$. Choosing $f_{\bX'}$ as a capacity-achieving distribution {in $\Omega(\beta')$}, the right-hand side of \eqref{ixy-lb} becomes $(1-\epsilon)\new{\Ci}(\beta')$. Thus,
\eqlab{eq:c-beta}{
\new{\Ci}(\beta) &\ge I(\bX;\bY) \notag\\
  &\ge (1-\epsilon) \new{\Ci}(\beta')
  .}
If now $\new{\Ci}(\beta) < \new{\Ci}(\beta')$, then \eqref{eq:c-beta} would yield a contradiction in the range $0<\epsilon<1-\new{\Ci}(\beta)/\new{\Ci}(\beta')$. Hence, $\new{\Ci}(\beta) \ge \new{\Ci}(\beta')$}.
\end{IEEEproof}

Intuitively, an input distribution with a nondecreasing mutual information for a given channel can be constructed by combining two parts, a high-probability part at a moderate cost, which does not vary much as the overall average cost is changed, and a low-probability part, a ``satellite,'' which absorbs the whole increase in average cost by moving away from the other part \cite{agrell12, hager12}. {As $\epsilon \rightarrow 0$, the input distribution $f_{\bX}$ becomes more and more like $f_{\bX'}$, while the {average cost $\E[b(\bx)]$} remains at $\beta$} {because the lower cost $\beta' < \beta$ is balanced by another cost $\beta'' \gg \beta$.}

\new{
The nondecreasing nature of the information capacity can be exploited to establish a channel coding theorem with equality cost constraint as follows. Even if Shannon and other information theorists may have been aware of the theorem, we have not seen it in print.

\begin{theorem}[Coding Theorem]
For any static point-to-point channel $f_{\bY|\bX}$,
\eqlab{eq:codingtheorem}{
  \C(\beta) = \sup_{f_\bX \in \Omega(\beta)} I(\bX;\bY)
.}
\end{theorem}

\begin{IEEEproof}
The theorem will be proved in two steps. First, it is shown that the operational channel capacities \eqref{eq:cprimedef} and \eqref{eq:cdef} are the same and second, that the information capacity $\sup I(\bX;\bY)$ is the same regardless of whether the optimization is over $\tilde{\Omega}(\beta)$ or $\Omega(\beta)$. The theorem then follows from the regular channel coding theorem \eqref{eq:cprime}.

For the first step, we use the relation
\eqlab{eq:Mineq}{
M(N,p,\beta) \le \tilde{M}(N,p,\beta) \le M(N+1,p,\beta)
,}
where the first inequality is trivial from the definitions of $M$ and $\tilde{M}$, whereas the second was proved by Shannon \cite[pp.~649--651]{shannon59}, \cite[eq.~(195)]{polyanskiy10}, who added an ($N+1$)th symbol to every codeword in a codebook $\C$ with codeword length $N$ and codeword cost at most $\beta$, to obtain a codebook $\tilde{\C}$ with codeword length $N+1$ and cost exactly $\beta$. Taking the logarithm of all three parts of \eqref{eq:Mineq}, dividing by $N$, and letting $N \rightarrow\infty$ proves via \eqref{eq:cprimedef} and \eqref{eq:cdef} that $C(\beta) \le \tilde{C}(\beta) \le C(\beta)$, in other words $C(\beta) = \tilde{C}(\beta)$.

For the second step, the information capacity \eqref{eq:cprime} is written as
\eqlab{eq:cprime2}{
\tilde{C}(\beta) = \sup_{\beta' \le \beta} \Ci(\beta')
,
}
which by Theorem \ref{th:main} is equal to $\Ci(\beta)$. Combining the two steps, $C(\beta) = \tilde{C}(\beta) = \Ci(\beta)$ and \eqref{eq:codingtheorem} follows.
\end{IEEEproof}

This means not only that a channel coding theorem holds for an equal-cost constraint but also that the two channel capacities \eqref{eq:cprimedef} and \eqref{eq:cdef}
}
are equivalent\new{.} The cost-limited channel capacity $\tilde{C}(\beta)$ is achieved by an input distribution $f_\bX$ for which the cost equals the maximum allowed value $\beta$. A practical interpretation is that when designing a capacity-achieving code for a nonlinear channel, it suffices to consider only codes for which all codewords have the same cost $\beta$.

\section{Interference Channels}

We consider a discrete-time, memoryless interference channel with $k$ users, each with the purpose of transmitting a message from a transmitter to a receiver \cite[Ch.~6]{elgamal11}, for example an optical WDM system. The input and output are denoted by $\bX_{\!i}$ and $\bY_{\!i}$, resp., for $i=1,\ldots,k$. The $i$th receiver attempts to recover $\bX_{\!i}$ based on $\bY_{\!i}$, without knowledge of $\bY_{\!j}$ for $j \ne i$. The statistics of the received vectors is given by the conditional distribution $f_{\bY_{\!1},\ldots,\bY_{\!k} | \bX_{\!1},\ldots,\bX_{\!k}}$, which does not change with the cost. Independent data is transmitted by each user,
and the joint {\emph{a priori}} input distribution $f_{\bX_{\!1},\ldots,\bX_{\!k}}$ is therefore equal to the product of the marginal distributions $f_{\bX_{\!1}}\cdots f_{\bX_{\!k}}$. All input distributions $f_{\bX_{\!i}}$ are known to all users. From the viewpoint of user $i$, all interfering input symbols $\bX_{\!j}$ for $j \ne i$ are assumed to be independent between channel uses. This assumption, which is conventional in optical communications, is valid if the codebook of user $j$ is not known to user $i$ or if user $j$ transmits uncoded data.

Three scenarios, or \emph{behavioral models} \cite{agrell13ofc}, are considered in the following subsections. \new{The aim in the first two scenarios} is to determine the maximum achievable rate of the primary user, referred to as user 1, while treating the signals from the other users $\bX_{\!2},\ldots,\bX_{\!k}$ as (nonlinear) noise. The received vectors $\bY_{\!2},\ldots,\bY_{\!k}$ are unknown at receiver 1 and the channel can be represented by the conditional distribution $f_{\bY_{\!1} | \bX_{\!1},\ldots,\bX_{\!k}}$. The third and last scenario represents joint optimization of $f_{\bX_{\!1}},\ldots,f_{\bX_{\!k}}$, considering the full interference channel model $f_{\bY_{\!1},\ldots,\bY_{\!k} | \bX_{\!1},\ldots,\bX_{\!k}}$. The point-to-point case extends straightforwardly to the first and third case (Sec.~\ref{sec:fixed} and \ref{sec:joint}), whereas the second case requires a somewhat more elaborate treatment (Sec.~\ref{sec:adaptive-dist}). There also exist behavioral models, not treated in this paper, for which the capacity is not monotonic \cite{agrell13ofc}.

The following lemma about conditional mutual information will be useful in Sec.~\ref{sec:adaptive-dist}.
\begin{lemma} \label{lem:i-diff}
For any $\bX$ and $\bY$, and any discrete $\bZ$,
\eq{
|I(\bX;\bY)-I(\bX;\bY|\bZ)| \le H(\bZ)
.}
\end{lemma}

\begin{IEEEproof}
By the chain rule for mutual information,
\eq{
I(\bX;\bY,\bZ) &= I(\bX;\bY) + I(\bX;\bZ|\bY), \\
I(\bX;\bY,\bZ) &= I(\bX;\bZ) + I(\bX;\bY|\bZ)
.}
Eliminating $I(\bX;\bY,\bZ)$ and rearranging terms,
\eq{
|I(\bX;\bY)-I(\bX;\bY|\bZ)| = |I(\bX;\bZ)-I(\bX;\bZ|\bY)|
.}
Since $\bZ$ is discrete by assumption, the right-hand side can be upperbounded using
\eq{
  0 &\le I(\bX;\bZ) \le H(\bZ), \\
  0 &\le I(\bX;\bZ|\bY) \le H(\bZ|\bY) \le H(\bZ),
}
which completes the proof.
\end{IEEEproof}

\subsection{Fixed Interference Distributions} \label{sec:fixed}

Suppose that the {input} distributions $f_{\bX_{\!2}},\ldots, f_{\bX_{\!k}}$ are fixed and do not change even if $f_{\bX_{\!1}}$ would change. From the viewpoint of the primary user, the interference caused by the other users can be included in the channel model. Since the conditional distribution $f_{\bY_{\!1} | \bX_{\!1}}$ in this case depends on the distributions $f_{\bX_{\!2}},\ldots,_{\bX_{\!k}}$, but not on $f_{\bX_{\!1}}$, Theorem \ref{th:main} applies and the capacity--cost function for the primary user is nondecreasing. {This scenario was called \emph{behavioral model (a)} in \cite{agrell13ofc}.}

\subsection{\new{Equal Distributions}}\label{sec:adaptive-dist}

\new{In this section and the next, we consider scenarios where the distributions of all users are governed by a \emph{network controller,} by assigning modulation formats and power levels to all users. While the next section discusses the case of joint optimization over all user distributions, we assume in this section that}
all users apply the same input distribution \new{$f_\bX$}, or linearly rescaled versions thereof.
\new{Hence, the joint distribution of the users is}
\eqlab{eq:rescaled-distribution}{
\new{f_{\bX_{\!1}, \ldots, \bX_{\!k}}(\bx_1,\ldots,\bx_k) = \prod_{i=1}^k \alpha_i^n f_{\bX}(\alpha_i \bx_i)
},}
for some given constants $\new{\alpha_1},\ldots,\alpha_k$.

An important special case is $\new{\alpha_1} = \cdots = \alpha_k = 1$, which makes all \new{marginal} distributions $f_{\bX_{\!1}}, f_{\bX_{\!2}}, \ldots, f_{\bX_{\!k}}$ identical. {This special case, called \emph{behavioral model (c)} in \cite{agrell13ofc}, was studied in, e.g., \cite{freckmann09, essiambre10}. The power scaling via $\new{\alpha_1},\ldots,\alpha_k$ provides additional degrees of freedom.}

\new{
The network controller may wish to select the distribution $f_\bX$ such that the achievable rate of any single channel, say, channel 1, is maximized. Hence, we define the \emph{constrained information capacity} of the primary channel as
\eqlab{eq:c-primary}{
  C_1(\beta) \triangleq \sup I(\bX_{\!1};\bY_{\!1})
,}
where the supremum is taken over all distributions of the form \eqref{eq:rescaled-distribution}, with $f_{\bX} \in \Omega(\beta)$. Clearly, the mutual information $I(\bX_{\!1};\bY_{\!1})$ is an achievable rate for this channel, for any $f_\bX$, and hence $C_1(\beta)$ is an achievable rate. In the context of optical communications, $C_1$ was studied in \cite{freckmann09, essiambre10,mecozzi12}.
By analogy with the point-to-point channel, it might be tempting to interpret $C_1$ as the \emph{maximum} achievable rate under certain conditions; however, we believe that no such claims can be made without a precisely stated coding theorem for the interference channel, which is presently lacking.
}

\begin{theorem}\label{th:adaptive-distributions}
\new{The constrained information capacity $C_1(\beta)$ is a nondecreasing function of $\beta>0$}, for any interference channel $f_{\bY_{\!1} | \bX_{\!1},\bX_{\!2},\ldots,\bX_{\!k}}$.
\end{theorem}

\begin{IEEEproof}
\new{Let $f_{\bX'} \in \Omega(\beta')$ be a distribution%
\footnote{Or, more precisely, a sequence of distributions.}
for which the supremum in \eqref{eq:c-primary} is attained, i.e., $\new{C_1}(\beta') = I(\bX_{\!1};\bY_{\!1})$}, at some cost $\beta' \ge 0$. We will show that $\new{C_1}(\beta) \ge \new{C_1}(\beta')$ for any $\beta \ge \beta'$.

For any given $\beta\ge\beta'$ and $0<\epsilon\le 1$, let 
\eq{
\beta'' \triangleq \beta'+\frac{\beta-\beta'}{\epsilon}
}
and let $f_{\bX''}$ be any distribution over $\X$ with $\E[b(\bX'')] = \beta''$. We now define a time-sharing random vector $\bX$ given an auxiliary binary random variable $Q$ such that
\eqlab{eq:timesharing}{
  \bX \triangleq \begin{cases}
    \bX', & Q = 0, \\
    \bX'', & Q = 1,
  \end{cases}
}
where $\Pr\{Q = 1\} = \epsilon$. This vector satisfies
\eq{
  \E[b(\bX)] &= (1-\epsilon)\E[b(\bX')]+\epsilon \E[b(\bX'')] \notag\\
  & = (1-\epsilon)\beta' + \epsilon \beta'' \notag\\
  & = \beta
  .}

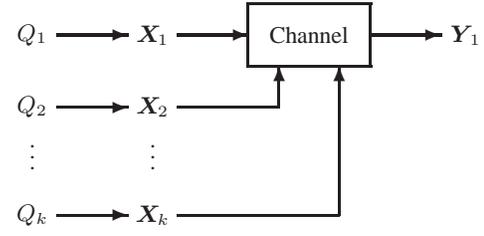
\begin{figure}
\begin{center}
\setlength{\unitlength}{.8mm} %
\small %
  \begin{picture}(75,38)(8,8)			%
    \thicklines
      \put(10,40){\makebox(0,0){$Q_1$}}			%
      \put(10,28){\makebox(0,0){$Q_2$}}			%
      \put(10,21){\makebox(0,0){$\vdots$}}			%
      \put(10,10){\makebox(0,0){$Q_k$}}			%
      \put(14,40){\vector(1,0){12}}		%
      \put(14,28){\vector(1,0){12}}		%
      \put(14,10){\vector(1,0){12}}		%
      \put(30,40){\makebox(0,0){$\bX_{\!1}$}}			%
      \put(30,28){\makebox(0,0){$\bX_{\!2}$}}			%
      \put(30,21){\makebox(0,0){$\vdots$}}			%
      \put(30,10){\makebox(0,0){$\bX_{\!k}$}}			%
      \put(34,40){\vector(1,0){12}}		%
      \put(34,28){\line(1,0){17}}		%
	\put(51,28){\vector(0,1){7}}		%
      \put(34,10){\line(1,0){27}}		%
	\put(61,10){\vector(0,1){25}}		%
      \put(46,35){\framebox(20,10){Channel}}%
      \put(66,40){\vector(1,0){12}}		%
      \put(82,40){\makebox(0,0){$\bY_{\!1}$}}			%
  \end{picture}
\caption{An interference channel with time-sharing inputs, analyzed in Sec.~\ref{sec:adaptive-dist}. The primary channel $\bX_{\!1}\rightarrow \bY_{\!1}$ is affected by interference from the other inputs $\bX_{\!2},\ldots,\bX_{\!k}${, which are all independent}.}
\label{fig:interference-channel}
\end{center}
\end{figure}

As illustrated in Fig.~\ref{fig:interference-channel}, the \new{joint distribution $f_{\bX_{\!1}, \ldots, \bX_{\!k}}$ is generated by an analogous $k$-fold} time-sharing method, using the auxiliary variables $\new{Q_1},\ldots,Q_k$. These variables have the same distribution as \new{$Q$} and are independent. They control the \new{input symbols} $\new{\bX_{\!1}},\ldots,\bX_{\!k}$ such that $\bX_{\!i} = \bX'_{\!i}$ if $Q_i = 0$ and $\bX_{\!i} = \bX''_{\!i}$ if $Q_i = 1$, where
\eqlab{eq:fxp-rescaled}{
f_{\bX'_{\!i}}(\bx) &= \alpha_i^n f_{\new{\bX'}}(\alpha_i \bx), \\
f_{\bX''_{\!i}}(\bx) &= \alpha_i^n f_{\new{\bX''}}(\alpha_i \bx) \label{eq:fxpp-rescaled}
}
for $i=\new{1},\ldots,k$. Obviously, the time-sharing \new{symbols $\bX_{\!i}$ jointly follow} the desired distribution \eqref{eq:rescaled-distribution}.

The mutual information of the primary channel can be bounded as
\eqlab{eq:i-rescaled0}{
I(\bX_{\!1}{;}\bY_{\!1}) &\ge I(\bX_{\!1};\bY_{\!1} | Q_1) \\
&\ge I(\bX_{\!1};\bY_{\!1} | Q_1, Q_2,\ldots, Q_k) \notag\\
&{\quad{}- H(Q_2,\ldots,Q_k)} \label{eq:i-rescaled}
,}
where \eqref{eq:i-rescaled0} holds because $Q_1\rightarrow\bX_{\!1}\rightarrow\bY_{\!1}$ is a Markov chain and \eqref{eq:i-rescaled} follows by setting $\bZ = [Q_2,\ldots,Q_k]$ in Lemma~\ref{lem:i-diff}. The first term of the right-hand side of \eqref{eq:i-rescaled} can be bounded as
\eqlab{eq:i-rescaled-cond}{
I(\bX&_{\!1};\bY_{\!1} | Q_1, Q_2,\ldots, Q_k) \notag\\
  &= \sum_{(q_1,\ldots,q_k)\in\{0,1\}^k} \Pr\{Q_1=q_1,\ldots,Q_k=q_k\} \notag\\
    &\qquad\qquad\qquad\quad\;\;\cdot {I(\bX_{\!1};\bY_{\!1} | Q_1=q_1,\ldots,Q_k=q_k)} \notag\\
  &\ge \Pr\{Q_1=\cdots=Q_k=0\} \notag\\
    &\quad\cdot {I(\bX_{\!1};\bY_{\!1} | Q_1=\cdots=Q_k=0)} \notag\\
  &= (1-\epsilon)^k I(\bX_{\!1}';\bY_{\!1}') \notag\\
  &= (1-\epsilon)^k \new{C_1}(\beta')
. }
The second term of the right-hand side of \eqref{eq:i-rescaled} is
\eqlab{eq:h-rescaled}{
H(Q_2,\ldots,Q_k) &= \sum_{i=2}^{k} H(Q_i) \notag\\
  &= (k-1) H_2(\epsilon)
,}
where $H_2(u) \triangleq -u \log_2 u -(1-u) \log_2(1-u)$. Combining \eqref{eq:c-primary}, \eqref{eq:i-rescaled}, \eqref{eq:i-rescaled-cond}, and \eqref{eq:h-rescaled} yields
\eq{
\new{C_1}(\beta) &= \sup_{f_{\new{\bX}}\in\Omega(\beta)} I(\bX_{\!1};\bY_{\!1}) \notag\\
  &\ge \sup_{0<\epsilon\le 1} \left[ (1-\epsilon)^k \new{C_1}(\beta')-(k-1) H_2(\epsilon) \right] \notag\\
  &= \lim_{\epsilon\rightarrow 0} \left[ (1-\epsilon)^k \new{C_1}(\beta')-(k-1) H_2(\epsilon) \right] \notag\\
  &= \new{C_1}(\beta')
  ,}
which completes the proof.
\end{IEEEproof}

Intuitively, the proof relies on constructing a ``satellite distribution'' \cite{agrell12} for \new{$\bX$}, where the ``satellite,'' denoted by \new{$\bX''$} in \eqref{eq:timesharing}, carries a much higher cost than \new{$\bX'$} and occurs with lower probability.

\subsection{Joint Optimization} \label{sec:joint}

In the third and last scenario, we assume that the system includes a mechanism to optimize the transmission schemes of all users jointly, for example via a central network controller. {As in the previous two scenarios, the transmitters and receivers are still \emph{operated} separately, in the sense that the transmitters and receivers do not exchange information about their respective signals.}\footnote{{If data instead is jointly encoded over all transmitted signals $\bX_{\!1},\ldots,\bX_{\!k}$ and jointly decoded based on all received signals $\bY_{\!1},\ldots,\bY_{\!k}$, then the channel is equivalent to a high-dimensional point-to-point channel and Theorem~\ref{th:main} applies.}}

Let $R_i$ be an achievable rate for the transmitter--receiver pair $i=1,\ldots,k$ and let $\bR \triangleq (R_1,\ldots,R_k)$ be a vector of rates that can be \emph{simultaneously} achieved over the interference channel, with arbitrarily small error probability. The \emph{capacity region} $\cC(\bbeta)$, where $\bbeta\triangleq(\beta_1,\ldots,\beta_k)$, is defined as the closure of the set of all achievable rate vectors $\bR$ when every codeword used by user $i=1,\ldots,k$ has the exact cost $\beta_i${\cite[Sec.~4.1, 6.1]{elgamal11}}. While no analytical expression is known for the capacity region of general interference channels \cite[Ch.~6]{elgamal11}, {the monotonicity can be established via the following theorem.}

\begin{theorem} \label{th:joint}
Let $\bbeta=(\beta_1,\ldots,\beta_k)$ and $\bbeta'=(\beta'_1,\ldots,\beta'_k)$ be two cost vectors such that $\beta_i \ge \beta'_i \ge 0$ for $i=1,\ldots,k$. Then their capacity regions satisfy $\cC(\bbeta) \supseteq \cC(\bbeta')$.
\end{theorem}

\begin{IEEEproof}
Let, for any $0<\epsilon\le 1$,
\eq{
\bbeta'' \triangleq \bbeta'+\frac{\bbeta-\bbeta'}{\epsilon}
.}
Let $\bR'$ and $\bR''$ be achievable rate vectors at costs $\bbeta'$ and $\bbeta''$, resp. By time sharing{\cite[Sec.~15.3.3]{cover06}, \cite[Sec.~4.4]{elgamal11}}, the rate
\eq{
(1-\epsilon)\bR' + \epsilon \bR'' \ge (1-\epsilon)\bR'
}
is achievable at cost
\eq{
(1-\epsilon)\bbeta' + \epsilon \bbeta'' = \bbeta
.}
The capacity region $\cC(\bbeta)$ thus includes all rate vectors of the form $(1-\epsilon)\bR'$, where $\bR'$ is achievable at cost $\bbeta'$ and $\epsilon$ is an arbitrarily small positive number. Since the capacity region by definition is the \emph{closure} of all achievable rate vectors{\cite[Sec.~4.1, 6.1]{elgamal11}}, $\cC(\bbeta)$ also includes $\lim_{\epsilon\rightarrow 0} (1-\epsilon)\bR' = \bR'$. 
In conclusion, $\bR' \in \cC(\bbeta)$ for all $\bR' \in \cC(\bbeta')$, which implies $\cC(\bbeta) \supseteq \cC(\bbeta')$.
\end{IEEEproof}

The capacity region is a $k$-dimensional object, and it varies as a function of the $k$-dimensional vector {$\bbeta$. The following two corollaries exemplify how linear combinations of the achievable rates change when the cost is varied linearly.}

\begin{corollary}\label{cor:C1}
If the cost is varied along a line as
\eq{
\bbeta = \bbeta_0+\mu \bDelta
,}
where all components of $\bbeta_0$ and $\bDelta$ are nonnegative, then all achievable rates $R_1,\ldots,R_k$ are nondecreasing functions of $\mu\ge 0$, and the achievable sum rate $R_1+\cdots+R_k$ is also a nondecreasing function of $\mu\ge 0$.
\end{corollary}

\begin{corollary}\label{cor:C2}
If all transmitters obey the same cost constraint $\beta_1 = \cdots = \beta_k = \beta$, then all achievable rates $R_1,\ldots,R_k$ are nondecreasing functions of $\beta$.
\end{corollary}

\section{Numerical Example} %

In this section, examples are given for mutual information and channel capacity as functions of the transmit power, for a simple nonlinear channel. The studied channel is chosen {mainly} for its simplicity, because evaluating the channel capacity is numerically possible only for very low-dimensional, memoryless channels, which unfortunately excludes more realistic channel models.

\subsection{A Nonlinear Channel}%

We consider a very simple channel with nonlinear distortion and additive noise, represented as
\eqlab{eq:channel}{
Y = a(X)+Z
,}
where $X$ and $Y$ are the input and output of the channel, resp., {$\X=\Y=\R$,} $a(\cdot)$ is a given deterministic function, and $Z$ is white Gaussian noise with zero mean and variance $\sigma_Z^2$. For a given channel input $x$, the channel {law is given} by the conditional probability density function (pdf)
\eqlab{eq:fyx}{
f_{Y|X}(y|x) = \frac{1}{\sigma_Z} f_G \left( \frac{y-a(x)}{\sigma_Z} \right)\!
,}
where $f_G(x) \triangleq (1/\sqrt{2\pi}) \exp(-x^2/2)$ is the zero-mean, unit-variance Gaussian pdf.
Since $f_{Y|X}(y|x)$ is Gaussian for any $x$, the conditional entropy is \cite[Sec.~20]{shannon48}, \cite[Sec.~8.1]{cover06}
\eqlab{eq:hyx3}{
h(Y|X) = \frac{1}{2}\log_22\pi e \sigma_Z^2
.}
For a given input distribution $f_X$, the output distribution $f_Y$ is obtained by marginalizing the joint distribution $f_{X,Y}(x,y) = f_X(x) f_{Y|X}(y|x)$, and the mutual information is calculated as $I(X;Y) = h(Y)-h(Y|X)$.

\begin{figure}
\begin{center}
\psfrag{ax}{\footnotesize $a(x)$}
\psfrag{x}{\footnotesize $x$}
\includegraphics[width=\columnwidth]{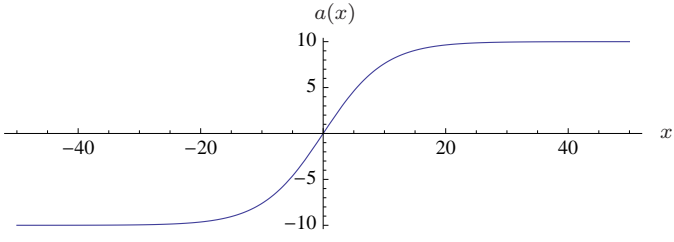}
\caption{A simple example of nonlinear distortion, given by \eqref{eq:tanh} for $\amax=10$. The channel is essentially linear for small $|x|$ and binary for large $|x|$.}
\label{fig:tanh}
\end{center}
\end{figure}

In this example, we select $a(x)$ in \eqref{eq:channel} as a smooth clipping function
\eqlab{eq:tanh}{
a(x) = \amax \tanh \left( \frac{x}{\amax} \right)\!
,}
where $\amax>0$ sets an upper bound on the output. {The hyperbolic tangent is commonly used to model nonlinear amplifiers \cite{sobhy96, abdelfattah06} and similar characteristics, albeit biased, can model a light-emitting diode in intensity-modulated optical systems \cite{elgala09, elgala10}.} If the instantaneous channel input $X$ has a sufficiently high magnitude compared with $\amax$, the channel is essentially binary. For $X$ close to zero, on the other hand, the channel approaches a linear AWGN channel.

The channel parameters are $\amax=10$ and $\sigma_Z=1$ throughout this section. The function $a(x)$ in \eqref{eq:tanh}, which represents the nonlinear part of the channel \eqref{eq:channel}, is shown in Fig.~\ref{fig:tanh}. {Since the channel law $f_{Y|X}(y|x)$ given by \eqref{eq:fyx} and \eqref{eq:tanh} depends on $x$, $y$, and $\amax$ but nothing else, the channel is static according to Definition~\ref{def:static} and Theorem~\ref{th:main} applies.}

\subsection{Mutual Information}%

\begin{figure}
\begin{center}
\psfrag{x}{\footnotesize $\p$}
\psfrag{y}{\footnotesize $I(X;Y)$ \new{[bits]}}
\psfrag{99}[][]{\scriptsize $0.1$}
\psfrag{0}[][]{\scriptsize $1$}
\psfrag{10}[][]{\scriptsize $10$}
\psfrag{20}[][]{\scriptsize $100$}
\psfrag{30}[][]{\scriptsize $1000$}
\psfrag{40}[][]{\scriptsize $10^4$}
\psfrag{50}[][]{\scriptsize $10^5$}
\includegraphics[width=\columnwidth]{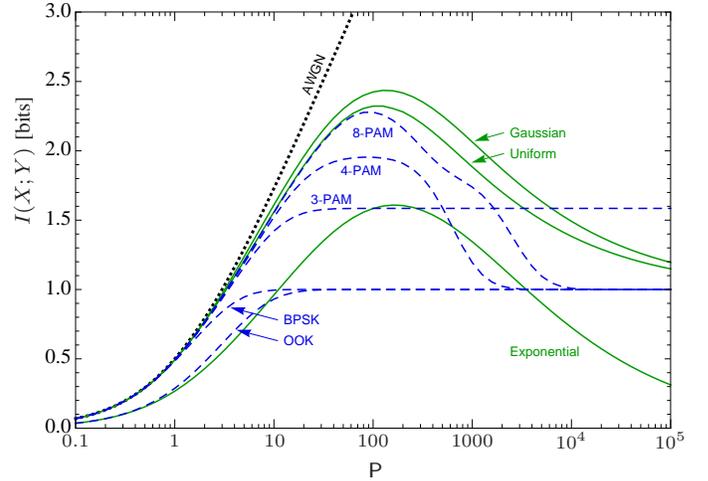}
\caption{Mutual information for the nonlinear channel in \eqref{eq:channel} with $\amax=10$ and $\sigma_Z=1$ with various continuous (solid) and discrete (dashed) input distributions. The discrete distributions have uniform probabilities and equal spacing. The AWGN channel capacity is included for reference (dotted).}
\label{fig:ixy}
\end{center}
\end{figure}

The mutual information $I(X;Y)$ is evaluated by numerical integration, as a function of the average transmit power $\p = \E[X^2]$. No optimization over input distributions is carried out. The input distribution $f_X(x)$ is constructed from a given unit-power distribution $g(x)$, rescaled to the desired power $\p$ as $f_X(x) = \alpha g(\alpha x)$, where $\alpha=1/\sqrt{\p}$. The results are presented in Fig.~\ref{fig:ixy} for three continuous input pdfs $f_X(x)$: zero-mean Gaussian, zero-mean uniform, and single-sided exponential, defined as, respectively,
\eq{
f_{X_1}(x) &= \frac{1}{\sqrt{\p}} f_G\left(\frac{x}{\sqrt{\p}}\right)\!, \\
f_{X_2}(x) &= \begin{cases}
  \frac{1}{2\sqrt{3\p}}, & -\sqrt{3\p} \le x \le \sqrt{3\p}, \\
  0, & \text{elsewhere},
\end{cases} \\
f_{X_3}(x) &= \begin{cases}
  \sqrt{\frac{2}{\p}}e^{-x \sqrt{2/\p}}, & x \ge 0, \\
  0, & x < 0.
\end{cases}
}

At asymptotically low power $\p$, the channel is effectively an AWGN channel. In this case, the mutual information is governed by the mean value of the input distribution, according to \cite{verdu02}. All zero-mean input distributions achieve approximately the same mutual information, which approaches the AWGN channel capacity. The asymptotic mutual information for the exponential distribution, whose mean is $\sqrt{\p/2}$, is half that achieved by zero-mean distributions.

The mutual information curves for all three input pdfs reach a peak around $\p=100$, when a large portion of the input samples still fall in the linear regime of the channel. When the transmit power $\p$ is further increased, the mutual information decreases towards a value slightly less than $1$ \new{bit} asymptotically for the zero-mean input pdfs and $0$ for the exponential input. The asymptotes are explained by the fact that at high enough power, almost all input samples fall in the nonlinear regime, where the channel behaves as a 1-bit noisy quantizer. {The same argument shows that for \emph{any} continuous distribution, the asymptotical mutual information is less than $1$ \new{bit}.}

Similar results for various discrete input distributions are also included in Fig.~\ref{fig:ixy}. The studied one-dimensional constellations are on--off keying (OOK), binary phase-shift keying (BPSK), and $m$-ary pulse amplitude modulation ($m$-PAM). The constellation points are equally spaced and the input samples $X$ are chosen uniformly from these constellations. The mutual information for $m$-PAM constellations with $m\ge 4$ exhibits the same kind of peak as the continuous distributions in Fig.~\ref{fig:ixy}; indeed, a uniform distribution over equally spaced $m$-PAM approaches the continuous uniform distribution as $m \rightarrow \infty$.

Similarly to the continuous case, the mutual information for zero-mean discrete input distributions approach the AWGN channel capacity as $\p\rightarrow 0$. Half this channel capacity is achieved by the OOK input, which has the same mean value $\sqrt{\p/2}$ as the exponential input above. The asymptotics when $\p\rightarrow\infty$ depends on whether {the input distribution includes a nonzero probability mass at $X=0$. If not,} the channel again acts like a 1-bit quantizer and the asymptotic mutual information is slightly less than 1. {For distributions with a probability mass at $X=0$,} here exemplified by $3$-PAM, the channel asymptotically approaches a ternary-output noisy channel {whose possible outputs are not only $Y=\pm \amax+Z$ but also $Y=0+Z$. Hence, for any input distribution (discrete, continuous, or mixed), the} mutual information is upperbounded by $\log_2 3 = 1.58$.

{In conclusion,} this particular channel has the property that the mutual information for any input distribution approaches a limit as $\p\rightarrow\infty$, and this limit is upperbounded by $\log_2 3$. It might seem tempting to conclude that the channel capacity, which is the supremum of all mutual information curves, would behave similarly. However, as we shall see in the next section, this conclusion is not correct, because the limit of a supremum is in general not equal to the supremum of a limit. Specifically, the asymptotical channel capacity is $\lim_{\p\rightarrow\infty} C(\p) = \lim_{\p\rightarrow\infty} \sup_{g} I(X;Y)$, which is not equal to $\sup_{g} \lim_{\p\rightarrow\infty} I(X;Y) \le \log_2 3$.

\subsection{Channel Capacity} \label{sec:channelcap} %

The standard method to calculate the channel capacity of a discrete memoryless channel is by the {\emph{Arimoto--Blahut algorithm}} \cite{arimoto72, blahut72}, \cite[Sec.~10.8]{cover06}, \cite[Ch.~9]{yeung08}. It has been extended to continuous-input, continuous-output channels in \cite{chang88} {and furthermore to cost-constrained inputs in \cite{dauwels05}. The idea in \cite{dauwels05} is to represent distributions} by lists of samples, so-called \emph{particles.} {A particle-based input distribution has the form}
\eqlab{eq:particle}{
f_X(x) = \sum_{i=1}^s w_i \delta(x-c_i)
,}
where $\delta(\cdot)$ is the Dirac delta function, $s$ is the number of particles, $\bc=(c_1,\ldots,c_s)$ are the particles, and $\bw=(w_1,\ldots,w_s)$ are the probabilities, or weights, associated with each particle. If $s$ is large enough, any distribution can be represented in the form \eqref{eq:particle} with arbitrarily small error. With this representation,
\eq{
f_Y(y) = \sum_{i=1}^s \frac{w_i}{\sigma_Z} f_G \left( \frac{y-a(c_i)}{\sigma_Z} \right)\!
,}
which yields $h(Y)$, and thereby $I(X;Y)$, by numerical integration.

{Since $h(Y|X)$ is constant, the capacity is obtained by maximizing $h(Y)$ subject to constraints on the total probability and power. This problem is in general nonconvex. In \cite{dauwels05}, the optimization is done by \emph{alternating optimization} \cite[Sec.~9.1]{yeung08}, first finding $\bw$ for a given $\bc$ using the Arimoto--Blahut algorithm and then finding $\bc$ for a given $\bw$ using a gradient search, and so on. Here, we apply gradient search techniques for both steps.}
The objective is to maximize the Lagrangian function
\eq{
L(\bc,\bw,\lambda_1,\lambda_2)
  &\triangleq h(Y)+\lambda_1\left( \sum_{i=1}^s w_i - 1 \right) \notag\\
  &\quad+\lambda_2\left( \sum_{i=1}^s w_i c_i^2 - \p \right)\!
,
}
where the Lagrange multipliers $\lambda_1$ and $\lambda_2$ are determined to maintain the constraints $\sum_i w_i = 1$ and $\sum_i w_i c_i^2 = \p$ during the optimization process. The gradients of $L$ with respect to $\bc$ and $\bw$ are calculated, and a steepest descent algorithm (or more accurately, ``steepest ascent'') is applied to maximize $L$. In each iteration, a step is taken in the direction of either of the two gradients.\footnote{Moving in the direction of the joint gradient turned out to be less efficient, because for small and large $\p$, the numerical values of $\bc$ and $\bw$ are not of the same order of magnitude.} The step size is determined using the \emph{golden section method}{\cite[Sec.~10.4]{lundgren10}}.
Several initial values $(\bc,\bw)$ were tried. {The number of particles $s$ was heuristically chosen by doubling its value until the obtained channel capacity changed by less than $0.01$. This convergence criterion was satisfied at $s=16$ in all cases.}

\begin{figure}
\begin{center}
\psfrag{x}{\footnotesize $\p$}
\psfrag{99}[][]{\scriptsize $0.1$}
\psfrag{0}[][]{\scriptsize $1$}
\psfrag{10}[][]{\scriptsize $10$}
\psfrag{20}[][]{\scriptsize $100$}
\psfrag{30}[][]{\scriptsize $1000$}
\psfrag{40}[][]{\scriptsize $10^4$}
\psfrag{50}[][]{\scriptsize $10^5$}
\psfrag{y}{\footnotesize $C(\p)$ \new{[bits]}}
\includegraphics[width=\columnwidth]{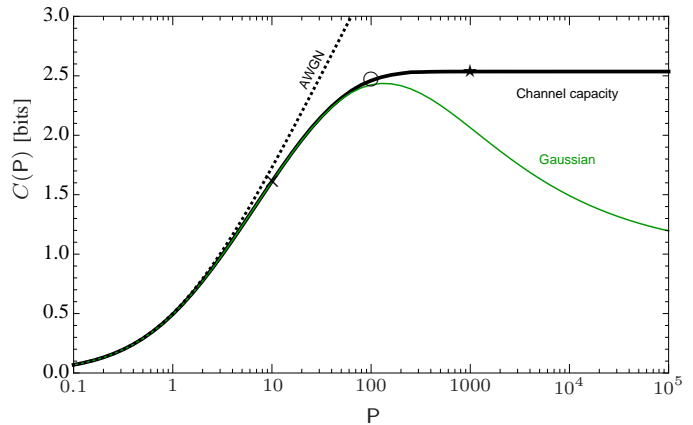}
\caption{Channel capacity for the same channel (thick solid), compared with the mutual information of the Gaussian distribution in Fig.~\ref{fig:ixy} (thin solid) and the AWGN channel capacity (dotted). Even though most mutual information curves decrease, the channel capacity does not, thus supporting Theorem~\ref{th:main}. The three markers refer to distributions in Fig.~\ref{fig:pmfs}.}
\label{fig:channelcap}
\end{center}
\end{figure}

The topography of $L$ as a function of $\bc$ and $\bw$ turned out to include vast flat fields, where a small step has little influence on $L$. This made the optimization numerically challenging. No suboptimal local maxima were found for the studied channel and constraints, although for nonlinear channels in general, the mutual information as a function of the input distribution may have multiple maxima.\footnote{An exception occurs when the constellation points $\bc$ are fixed and the only constraint is $\sum w_i=1$. In this special case, the mutual information is a concave function of $\bw$ for any channel{\cite[Sec.~2.7, 7.3]{cover06}} and there is thus a unique maximum.}

\begin{figure*}
\begin{center}
\psfrag{x}{\footnotesize $x$}
\psfrag{amax}{\footnotesize $\amax$}
\psfrag{pow10}{\small $\p=10$}
\psfrag{pow100}{\small $\p=100$}
\psfrag{pow1000}{\small $\p=1000$}
\includegraphics[width=\textwidth]{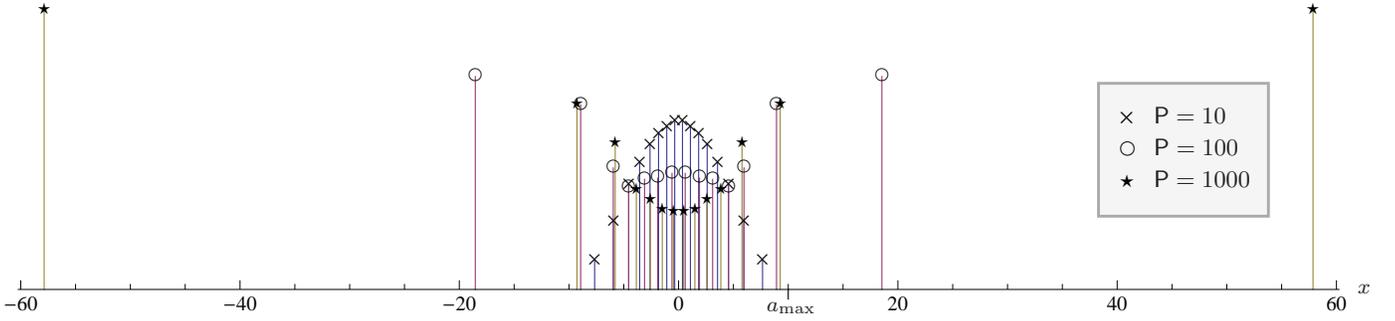}
\caption{Discrete approximations of the capacity-achieving input distributions for $\p=10$, $100$, and $1000$.}
\label{fig:pmfs}
\end{center}
\end{figure*}

This channel capacity, numerically obtained by the above method, is shown in Fig.~\ref{fig:channelcap} for the studied channel (thick solid curve). As promised by \new{Theorem \ref{th:main}}, the curve differs from most mutual information curves by not having a peak at any $\p$. The channel capacity follows the mutual information of the Gaussian distribution closely until around $\p=100$. However, while the Gaussian case attains its maximum mutual information $I(X;Y) = 2.44$ bits/symbol at $\p=130$ and then begins to decrease, the channel capacity continues to increase towards its asymptote $\lim_{\p\rightarrow \infty} C(\p) = 2.54$ bits/symbol. {The fact that the capacity curve rises somewhat over the peak and not only flattens out is encouraging for future work on capacity-achieving coding for more realistic nonlinear channel.}

This asymptotical channel capacity can be explained as follows. Define the random variable $A\triangleq a(X)$. Since $a(\cdot)$ is a continuous, strictly increasing function, there is a one-to-one mapping between $X \in (-\infty,\infty)$ and $A \in (-\amax,\amax)$. Thus $I(X;Y) = I(A;Y)$, where $Y=A+Z$. This represents a standard discrete-time AWGN channel whose input $A$ is subject to a peak power constraint. The capacity of a peak-power-constrained AWGN channel was bounded already in \cite[Sec.~25]{shannon48} and computed numerically in \cite{smith71}, where it was also shown that the capacity-achieving distribution is discrete. The asymptote in Fig.~\ref{fig:channelcap}, which is $2.54$ bits/symbol or, equivalently, $1.76$ nats/symbol, agrees perfectly with the amplitude-constrained capacity in \cite[Fig.~2]{smith71} for $\amax/\sigma_Z = 10$.

Some almost capacity-achieving input distributions are shown in Fig.~\ref{fig:pmfs}, numerically optimized as described above. For $\p=10$, the optimized discrete input distribution is essentially a nonuniformly sampled Gaussian pdf, and the obtained channel capacity, 1.61, has the same value as the mutual information of a continuous Gaussian pdf, shown in Fig.~\ref{fig:ixy}. For $\p=100$ and $1000$, the distribution is more uniform in the range where the channel behaves more or less linearly, which for this channel is approximately at $-\amax/2<x<\amax/2$, with some high-power outliers in the nonlinear range $|x| > \amax$. In all cases, increasing the number of particles $s$ from what is shown in Fig.~\ref{fig:pmfs} does not increase the mutual information significantly, from which we infer that these discrete input distributions perform practically as well as the best discrete or continuous input distributions for this channel.

Although the capacity-achieving distributions would look quite different for other types of nonlinear channels, a general observation can be made from Fig.~\ref{fig:pmfs}: Even at high average power, the input should consist of samples with moderate power, for which the channel is good, most of the time. The high average power is achieved by a single particle having a very large power; thus, the capacity-achieving distribution is a satellite distribution \cite{agrell12}. {This single particle, or satellite, corresponds to $\bX''$ and $\bX_{\!1}''$ in the proofs of Theorems \ref{th:main} and \ref{th:adaptive-distributions}, resp., which as $\epsilon \rightarrow 0$ have high cost (power) and low probability.}

\section{Summary and Conclusions}

It was proved that the channel capacity is a nondecreasing function of a cost (such as transmit power) in the following cases{:}
\begin{itemize}
\item Point-to-point memoryless vector channels $f_{\bY|\bX}$ that do not change with the input distribution $f_{\bX}$.
\item Interference channels where all users, except the one of interest, transmit data from fixed input distributions.
\item Interference channels where all users transmit data from the same (optimized) distribution.
\item Interference channels where the distributions of all users are optimized jointly.
\end{itemize}
The mutual information may be decreasing with cost in all these cases, but not the channel capacity in Shannon's sense.

In contrast, there are numerous examples in the literature where the channel capacity has a peak at a certain cost, after which it decreases towards zero{\cite{splett93, mitra01, stark01, tang01, green02, narimanov02, wegener04, kahn04, ho05, djordjevic05, tang06, taghavi06, essiambre08, freckmann09, essiambre10, djordjevic10, ellis10, smith10, goebel11, killey11, bosco11, mecozzi12}}. These examples all pertain to one of the following cases:
\begin{itemize}
\item Point-to-point channels that change depending on the transmitter settings, typically as a function of the transmit power \cite{agrell13ecoc}.
\item Interference channels where the transmission scheme of one user (the one of interest) is optimized while the other users satisfy the same power constraint by pure amplification \cite{agrell13ofc}.
\end{itemize}

 {A practical interpretation is that when designing codes for nonlinear channels under the constraint of a maximum average power, it suffices to consider codes in which all codewords satisfy the power constraint with equality. This is in contrast to previous works in optical communications, which often assumed the existence of an optimal (finite) power. Further research is needed to show whether the new approach is just a way to achieve the same rates as before at a higher power, or if it may lead to significantly increased achievable rates.}

\section*{Acknowledgments}
The author is indebted to 
A.~Alvarado,
P.~Bayvel, %
L.~Beygi,
G.~Durisi,
T.~Eriksson,
R.-J.~Essiambre, %
C.~H\"ager, %
M.~Karlsson,
J.~Karout,
G.~Kramer, %
F.~R. Kschischang, %
E.~Telatar, %
and S.~K. Turitsyn
for helpful discussions, criticism, and comments on early versions of this manuscript. {The presentation was greatly improved thanks to detailed feedback from four insightful reviewers.}

\balance

\end{document}